\documentclass[12pt,twoside]{article}   
\usepackage{latexsym} 
\usepackage{mathptm}
\usepackage{amsmath}
\usepackage{amssymb}
\usepackage{graphicx}
\def\beq{\begin{equation}}
\def\eeq{\end{equation}}
\def\beqn{\begin{eqnarray}}
\def\eeqn{\end{eqnarray}}
\begin{document}
 
\title{The Box-Problem in Deformed Special Relativity}
\author{Sabine Hossenfelder \thanks{hossi@nordita.org}\\
{\footnotesize{\sl Nordita, Roslagstullsbacken 23, 106 91 Stockholm, Sweden}}}
\date{}
\maketitle
 
\vspace*{-1cm}
\begin{abstract}
We examine the transformation of particle trajectories in models with deformations of Special Relativity that
have an energy-dependent and observer-independent speed of light. These transformations necessarily imply
that the notion of what constitutes the same space-time event becomes dependent on the observer's inertial frame. To
preserve observer-independence, the such arising nonlocality should not be in conflict with our
knowledge of particle interactions. This requirement allows us to derive strong bounds on 
deformations of Special Relativity and rule out a modification to first order in energy over the Planck mass.
\end{abstract}

\section{Introduction}

It is generally believed that the Planck mass $m_{\rm Pl}$
is of special significance. As the 
scale where effects of the yet-to-be-found theory of quantum gravity are expected to become important, it has been argued the energy associated to the 
Planck mass should have an observer-independent meaning. Lorentz-transformations however do not 
leave any finite energy invariant. Thus, the requirement of assigning an observer-independent
meaning to the Planck mass seems to necessitate a modification of Special Relativity and
a new sort of Lorentz-transformations. This modification of Special Relativity, 
which does not introduce a preferred frame but instead postulates the Planck mass as an observer-independent invariant, 
has become known as ``Deformed
Special Relativity'' (DSR) \cite{AmelinoCamelia:2000ge,KowalskiGlikman:2001gp,AmelinoCamelia:2002wr,Magueijo:2002am}.

The deformed Lorentz-transformations that leave the Planck mass 
invariant under boosts can be explicitly constructed. There are infinitely
many of such deformations, and they generically result in a modified dispersion relation
and an energy-dependent speed of light \cite{Hossenfelder:2005ed}. In the low energy limit, this energy-dependent
speed of light coincides
with the speed that we have measured. Depending on the sort of
deformation, the speed of light can increase, decrease, or remain constant
with energy. We will here examine the case where it is not constant.

These deformations of Special Relativity have recently obtained increased
attention since measurements of gamma ray bursts observed by the Fermi Space Telescope have
now reached a precision high enough to test a modification
in the speed of light to first order in the energy over the Planck mass \cite{Science,Nature,AmelinoCamelia:2009pg}. 
While
such modifications could also be caused by an actual breaking of Lorentz-invariance
that introduces a preferred frame, models that break Lorentz-invariance 
are subject to many other constraints already \cite{Maccione:2007yc}. This makes {\sc DSR} the prime 
candidate for an energy dependent speed of light. We will here argue 
that {\sc DSR} necessitates violations of locality that put much stronger
bounds on an energy-dependent speed of light already than the recent measurements of gamma ray bursts. 

This paper is organized as follows. In the next
section we will study a thought-experiment that lays out the basic problem that
an energy-dependent but observer-indepen\-dent speed of light renders locality a 
frame-dependent notion.
In section \ref{2.0}, we will transfer this thought-experiment into a
realistic setting. We will show that, with a first order modification of the
speed of light, the violations of locality would be within current measurement
precision and thus cannot be dismissed on grounds of practical impossibility of
detection. In section \ref{2.1} we will consider a variant
of the setup that covers the case in which there is an additional enhanced quantum
mechanical uncertainty in {\sc DSR} and show that still the problem is within current
measurement precision. This then requires us to put bounds on the energy
dependence of the speed of light such that the previously studied effect
is not in conflict with already existing measurements. This will be done
in section \ref{bounds}. In section \ref{disc} we will consider some
alternative options to prevent these bounds 
but have to conclude that these
are all implausible. We use the convention $c=\hbar=1$.

\section{The Box-Problem, Version 1.0}
\label{1.0}

In the cases of DSR we will examine, the speed of light is a function of 
energy $\tilde c(E)$, such that this function is the same for all observers. 
Thus, in a different restframe
where $E$ was transformed into $E'$ under the deformed Lorentz-transformation, the
speed of light would be $\tilde c'(E') = \tilde c(E')$. In ordinary Special Relativity
it is only one speed, $\lim_{E\to 0} \tilde c(E) =1$, that is invariant 
under the Lorentz-transformations. This is a result of deriving Lorentz-transformations 
as the symmetry-group of Minkowski space and not an assumption for the derivation. It 
is thus puzzling how an energy-dependent speed of
light that takes different values can also be observer-independent.

The intuitive problem can be seen in the following scenario. Consider the
case in which the speed of light was decreasing monotonically and finally
reached zero when the energy equaled the Planck mass. Then, a photon with
$E=m_{\rm Pl}$ would be at rest. We put this photon inside a box. The 
box represents a classical, macroscopic, low-energy object, one for which
 modifications of Special or General Relativity are absent or at least
negligible. 

What does an observer moving relative to the
box with velocity $v$ see? He sees the box move with $-v$ relative to
him. The photon's energy in his restframe is also the Planck mass,
since it is an invariant of the deformed Lorentz-transformation. 
Consequently the photon is also at rest, and cannot remain inside the
box. Indeed, if the observer only waits long enough, the photon will be
arbitrarily far outside the box.

If we bring another particle
into the game, for example an electron, that in the restframe of the box 
interacts with the photon, then the moving observer will generically 
see the particles interact outside the box (except for the specifically
timed case in which the electron just meets the photon in the moment when the
photon is also in the box). The different transformation behavior
of the world-lines of the box and the photon thus results in an
observer-dependent notion of what constitutes `the same' spacetime
event. In contrast to the observer-dependence of `the same' moment in
time that one also has in Special Relativity, this concerns the
observer dependence of what happens at the same time {\sl and} the same place.
Since two straight, non-parallel lines always meet in one point, an
example requires at least three lines, rspt. three objects moving with constant
velocity, here the photon, the electron and the box. In one reference
frame they all meet in the same space-time point. In another reference
frame they do not. This poses significant challenges if one wants to
accommodate it in a local theory.

While this setting exemplifies the box-problem, it can be criticized on the
grounds that experimentalists do not have many reasons to worry about
particles with energies of $10^{19}$~GeV. We will thus in the next section
study an actually observable situation. This will be a more
complicated setup, but the underlying cause of the problem remains
the same. It is the requirement that the speed of photons changes
with energy but changes in an observer-independent way that forces
upon us that the world-lines of particles transform differently depending
on the particle's energy. This then has the effect that the question
what constitutes `the same' spacetime event becomes observer-dependent,
which can run into conflict with observations that have confirmed the locality
of particle interactions to high precision.

\section{The Box-Problem, Version 2.0}
\label{2.0}

Consider a
gamma ray burst ({\sc GRB}) at distance $L \approx 4$~Gpc that, for simplicity, has no motion
relative to the Earth. This source emits a
photon with $E_\gamma \approx 10$~GeV, such that it arrives in the Earth
restframe at $(0,0)$ inside a detector. Together with the 10 GeV 
photon there is a low-energetic reference photon emitted. The energy of that photon can be as
low as wanted. 

In the {\sc DSR} scenario we are considering the dispersion relation of photons is modified to
\beqn
E^2 = p^2 + 2 \alpha \frac{E^3}{m_{\rm Pl}} + {\mbox{higher order}}\quad,
\label{disp}
\eeqn
and the phase velocity 
depends on the photons' energy. To first order
\beqn
\tilde c(E) \approx \left( 1 + \alpha \frac{E}{m_{\rm Pl}} \right) + {\cal O} \left(\frac{E^2}{m^2_{\rm Pl}} \right) \quad.
\label{alpha}
\label{cofe}
\eeqn
where we will neglect corrections of order higher than $E_\gamma/m_{\rm Pl}$ in the following, and set $\alpha = -1$, in
which case the speed of light decreases with increasing energy.

The important point is that Eq. (\ref{disp}) 
and (\ref{cofe}) are supposed to be observer independent, such that these relations have the same form in every reference frame. This then requires the non-linear, deformed 
Lorentz-transformations in momentum space. These transformations depend on
the form of the modified dispersion relation. We will however here work in an approximation and only need to know that
the Lorentz-transformations receive to lowest order a correction in $E/m_{\rm Pl}$. 

The 
higher energetic photon is slowed down and arrives
later than the lower energetic one. One has for the difference $\Delta T$ 
between the arrival times of the high and low energetic photon 
\beqn
\Delta T = L \left(\frac{1}{\tilde c (E_\gamma)} - 1 \right) = L \frac{E_\gamma}{m_{{\rm Pl}}} + {\cal O} \left(\frac{E_\gamma^2}{m^2_{\rm Pl}} \right) \quad. \label{DeltaT}
\eeqn
With 4 Gpc $\approx 10^{26}$~m, $E_\gamma \approx 10^{-18} m_{{\rm Pl}}$, the delay is of the order 1 second, take or give
an order of magnitude. Strictly speaking, this
equation should take into account the cosmological redshift since the photon propagates
in a time-dependent background. However, for our purposes of estimating the effects it will suffice to 
consider a static background, since using the proper General Relativistic expression does not change the
result by more than an order of magnitude \cite{Ellis:2002in,Jacob:2008bw}.

We further consider an electron at $E_e \approx 10$~MeV emitted from a source in
the detector's vicinity such that it arrives together with the high energetic photon at $(0,0)$ inside
the detector.  The source
can be as close as wanted, but to make a realistic setup it should be at least of 
the order $1$~m away from the detection point. 
The low energetic photon leaves the GRB together with the high energetic photon 
at $(x_e, t_e) = (-L, - L/\tilde c)$. It arrives in the detector box at $(x_a, t_a)=(0,L(1-1/\tilde c))$,
by $-t_a$ earlier than the electron.
We have chosen the emission time such that $-t_a=\Delta T$, and the electron arrives
with the same delay after the low energetic photon as the high energetic photon. 

With an energy of 10~MeV, the
electron is relativistic already, but any possible energy-dependent {\sc DSR} effect is at least 3
orders of magnitude smaller than that of the photon, and due to the electron's nearby emission the effects cannot accumulate over a long distance. The electron's velocity is
\beqn
v_e \approx  \left( 1 - \frac{1}{2}\frac{m_e^2}{E_e^2}\right) \approx \left( 1 - 10^{-3} \right) +O(E^2_e/m_{\rm Pl}^2)\quad.
\eeqn 

\begin{figure}[ht]
\includegraphics[width=11.5cm]{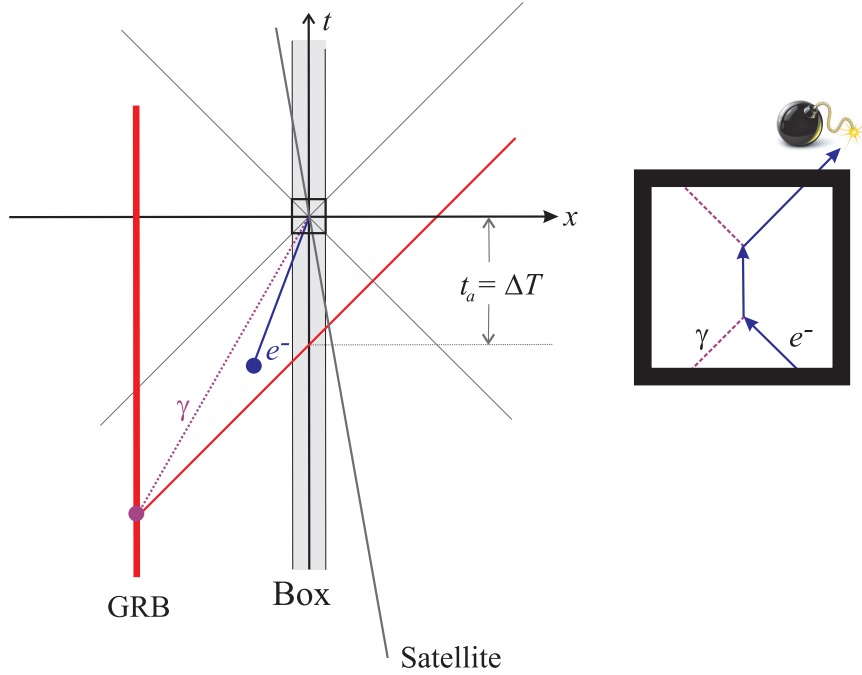}
\caption{{\small Labframe. The gamma ray burst (thick red line) is in rest with the detector (grey shaded area). It emits
at the same time one low energetic photon (thin red line) and one high energetic photon (dotted purple line) that is
slowed down due to the energy-dependent speed of light. From a source close to the detector, there is an
electron emitted (blue line) that meets the low energetic photon in the detector. The electron scatters on the
photon, changes momentum and triggers a bomb. A satellite flies by towards
the gamma ray burst and crosses the detector just when the photon also meets the electron. The thin grey lines 
depict the light-cone in the low energy limit.}}
\label{1}
\end{figure}

Inside the detector at $x=0$ the photon scatters off the electron. The photon changes 
the momentum of the electron, which triggers a bomb and the lab blows up. 
That is of course completely irrelevant. It only matters that the elementary scattering 
process can cause an irreversible and macroscopic change. 
This setup is depicted in Fig. \ref{1}.

Also in the picture is a satellite moving relative to the Earth restframe (thick grey
line in Fig. \ref{1}). From that satellite, a team of physicists observes and tries to describe the 
processes in the lab. The satellite crosses the lab just when the bomb blows off at $(0,0)$. That's somewhat 
of a stretch, but let's not overdo it with the realism. The typical speed of a 
satellite in Earth orbit is $v_S= -10$ km/s, or, in units of $c$, $v_S \approx - 3 \times 10^{-5}$, and
the gamma factor is approximately $\gamma_S \approx 1+10^{-9}$ for the relative motion between lab and satellite.   
Of course the satellite is bound in the gravitational field of the Earth and
not on a constant boost, but on the timescales that matter for the following
this is not relevant. Alternatively, replace Earth by a space station with
negligible gravitational field.

\begin{figure}[ht]
\includegraphics[width=11.5cm]{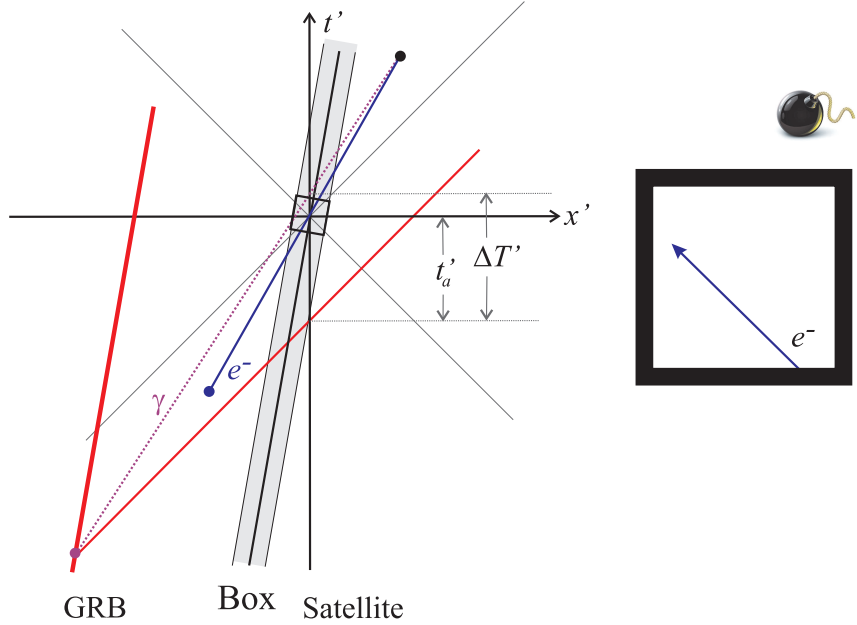}
\caption{{\small The same scenario as in Fig. \ref{1} as seen from the satellite restframe. The
gamma ray burst (thick red line) now moves to the right, and emits the low energetic photon (thin
red line) and the high energetic photon (dotted purple line) at slightly blueshifted energies. 
The high energetic photon is slowed down even more, misses the electron and the bomb is not triggered.}}
\label{2}
\end{figure}

Now let us look at the same scenario from the satellite restframe, shown
in Fig \ref{2}. We will denote the coordinates of that restframe with $(x',t')$. 
The satellite is moving towards the {\sc GRB}, thus the electron's
and photons' energies are blueshifted. We have
\beqn
E'_\gamma &=& {\sqrt\frac{1-v_S}{1+v_S}} E_\gamma + {\cal O} \left(\frac{E^2_\gamma}{m^2_{{\rm Pl}}} \right)  \quad,
\eeqn
and the energy of the very low energetic photon remains very low energetic.
The low-energetic photon crosses the satellite at $(x,t) = (L(1/\tilde c(E_\gamma)-1)/(1-1/v_S),L(1/\tilde c(E_\gamma)-1)/(v_S-1))$.
In the satellite frame the time passing between the arrival of the low energetic
reference photon and the electron at $x'=0$ is
\beqn
t'_a = \frac{L}{\gamma_S} \frac{1/\tilde c(E_\gamma) -1}{1 - v_S}  \quad. 
\eeqn
(Note that this is not the Lorentz-transformation of $t_a$, as becomes clear from the figures.) 
The formulation of {\sc DSR} in position space has been under debate.
It has been argued that the space-time metric should become energy-dependent 
\cite{Magueijo:2002xx,Kimberly:2003hp,Galan:2004st,Amelino-Camelia:2005ne}, and in
\cite{Hossenfelder:2006rr} it was shown that keeping the energy-dependent speed of light observer-independent
forces one to accept also the transformations in position-space become dependent
on an external parameter characterizing the particle (for example its energy), 
though the interpretation remains unclear. Thus, to keep track of assumptions made, let us point out 
that we are talking here about an observation
made on two low energetic particles from a very macroscopic, non-relativistic satellite. Even if there
was a {\sc DSR}-modification to the above transformation, it could come in here only through corrections of the 
order $E_e/m_{{\rm Pl}}$, and do so without this tiny contribution being able to add up over a long distance. 

With higher energy, the speed of the electron increases. The speed of the 
photon also changes but, and here is
the problem,  according to {\sc DSR} by assumption
the function $\tilde c$ is {\sl observer-independent}. In the satellite
frame one then has
\beqn
\tilde c (E_\gamma') = 1 - \frac{E'_\gamma}{m_{{\rm Pl}}} = 1 - {\sqrt\frac{1-v_S}{1+v_S}} \frac{E_\gamma}{m_{\rm Pl}} + {\cal O} \left(\frac{E_\gamma^2}{m^2_{\rm Pl}} \right) \quad,
\eeqn
and the distance the photons travel until they reach the satellite is
\beqn
L'= \gamma_S \left( v_S/\tilde c(E_\gamma) - 1 \right) L \quad. 
\eeqn
Thus, the time passing between the arrival of the reference photon and
the high energetic photon at the satellite is
\beqn
\Delta T' = \frac{E'}{m_{\rm Pl}} L' = \frac{1-v_S}{1+v_S} \Delta T +  {\cal O} \left(\frac{E_\gamma^2}{m^2_{{\rm Pl}}} \right) \quad. 
\label{dt}
\eeqn
Again, the question arises whether there could be some energy dependence in this transformation. Since  we are talking about passive transformations here, 
this creates an interpretational mess, 
but nevertheless we will discuss this possibility later in section \ref{disc}.
With the above, in the satellite frame the high energetic photon thus arrives later than the electron by 
\beqn
 \Delta T' - t'_a  = \left( \frac{1-v_S}{1+v_S} - \frac{1}{ \gamma_S(1-v_S)} \right)  \Delta T +  {\cal O} \left(\frac{E_\gamma^2}{m^2_{\rm Pl}} \right)  \quad.
\eeqn
Inserting $1/\gamma_S \approx 1 -1/2 v^2_S$ for $v_S\ll 1$, one finds
\beqn
\Delta T'  - t'_a \approx -3 \Delta T \left(v_S - \frac{1}{2} v_S^2 \right)  \approx 10^{-5} \Delta T \quad.
\eeqn
In the satellite frame, the low-energetic photon thus misses the electron by $\approx 10^{-5}$ seconds. Possible additional 
{\sc DSR} effects for the electron are negligible because of its low energy and short travel distance and
thus cannot save the day.
 
Now $10^{-5}$ seconds might
not appear much given the typical time resolution for detection of such particles is at best of the order milliseconds. However,
multiplied by the speed of light, the high energetic photon is still lagging behind as much as a kilometer when it arrives in the detector.
It only catches up with the electron at 
\beqn
x' = \frac{t'_a - \Delta T'}{1/\tilde{c}(E'_\gamma) - 1/v'_e}\approx  (t'_a - \Delta T') \frac{E^2_e}{m^2_e}  \approx 10^{5}~ {\rm m} \quad,
\eeqn
and thus safely outside the detector. The photon then cannot scatter off the electron in the detector, and
the electron cannot trigger the bomb to blow up the
lab. The physicists in the satellite are puzzled.

\section{The Box-Problem, Version 2.1}
\label{2.1}

An assumption we implicitly made in the previous section was 
that the quantum mechanical space- and time-uncertainties $\Delta t, \Delta x$ are not modified in {\sc DSR},
such that the GeV photon can be considered peaked to a $\Delta t$ smaller than the distance to the electron at arrival. For a distance of 
1 km, this is about 19 orders
of magnitude higher than $1/E_\gamma$ and thus an unproblematic assumption. 

Whether or not {\sc DSR} has  a modification of quantum mechanics is hard to say in absence 
of a formulation of the model in position space, so let us just examine the possibilities. There either
is a modification, or there is not. The previous section examined the case in which there is no modification. 
Here we will consider the case that
there was a modification of quantum mechanics. We will show that if the difference in arrival time in
the Earth frame $\Delta T$ was of the order seconds, this would either be incompatible with experiment, or
with observer independence. Later, we can use the experimental limits to obtain a on bound the possible
delay compatible with experiment.

The question whether or not the wave function spreads in {\sc DSR} depends on how one
interprets the modified dispersion relation. It is supposed to describe the
propagation of a particle in a background that displays quantum gravitational
effects. Yet the question is whether this modification should be understood as
one for a plane wave or for a localized superposition of plane waves already. In
the first case a wave-packet would experience enhanced dispersion, in the latter
case not. In the absence of a derivation, both interpretations seem plausible.

Let us point out that we are here talking about the dispersion during propagation
and the position uncertainty resulting from this and not a modification of the 
maximally possible localization itself. {\sc DSR} 
generically does not only have an energy-dependence of the speed of light, but also 
an energy-dependence of Planck's constant $\hbar$ \cite{Hossenfelder:2005ed}. This results in a generalized
uncertainty principle which in particular has the effect that particles with
momentum approaching the Planck scale have an increasing position uncertainty,
 as opposed to the limit on position
uncertainty monotonically decreasing with the ordinary Heisenberg relation. However, these
DSR corrections to $\hbar$ also go with powers of $E/m_{\rm Pl}$. This means that the maximally
possible localization of the 10 GeV photon at emission is affected, but to
an extend that is negligible. The relevant contribution to the uncertainty would
be the one stemming from the dispersion during propagation.

In case there is a modification caused
by a dispersion of the wave-packet, then the uncertainty of the slowed down, high energetic photon at arrival would 
be vastly larger than the maximal localization of the Heisenberg limit allows. If one starts with a Gaussian
wave-packet localized to a width of $\sigma_0$ at emission and tracks its spread with the modified dispersion
relation, one finds that to first order the now time-dependent width is
\beqn
\sigma(t) = \sigma_0 \sqrt{1 + \left( \frac{2 t}{m_{\rm Pl} \sigma^2_0}\right)^2}
\eeqn 

\begin{figure}[ht]
\includegraphics[width=11.5cm]{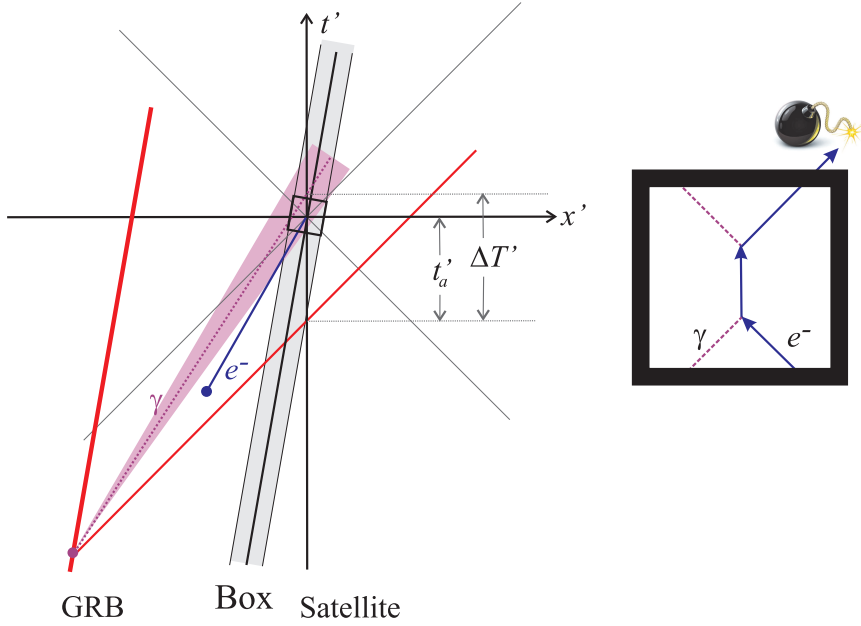}
\caption{{\small Satellite frame, with increased quantum uncertainty. The same scenario as in Fig. \ref{2} with added space and time 
uncertainty for the high energetic photon (purple area). The photon is smeared out all over the detector. It interacts with the electron and triggers the bomb without that interaction appearing nonlocal for the observer in the satellite.}
}
\label{4}
\end{figure}

If we start with a width of $\sigma_0 \approx 1/E_\gamma$, then for times $t \gg m_{\rm Pl} \sigma_0^2$ (which
amounts for the values we used to $t \gg 10^{-6}$~ seconds), one finds that the width is to first order 
$\sigma(t) \approx 2 t E_\gamma/m_{\rm Pl}$. Or, in other words,  
in the worst case the uncertainty of the wave-packet at arrival is about the same size as
the time delay $\Delta t \approx \Delta T$. 
In this case the photon 
at arrival would be smeared out over some hundred thousand
kilometers. A delay of $\Delta T$ with an uncertainty of $\Delta T$ is hard to detect, but it would also be  
impossible to find out whether or not the center of the wave-packet had been dislocated by a factor five orders of
magnitude smaller than the width of the wave-packet. This is sketched in Fig. \ref{4}. 

We recall however that the box-problem was caused by the unusual transformation behavior of $\Delta T$. To
entirely hide this behavior, the quantum mechanical uncertainty $\Delta t$ needs to be much larger than the delay 
$\Delta T - t_a$ in all restframes, such that it was practically unfeasible to ever detect a tiny difference
in probability with the photons we can receive, say, in the lifetime of the universe. We run into a problem when the
delay between the electron and the slow photon is about equal to or even smaller than the uncertainty of the slow photon. The two times $\Delta T$ and $t_a$ however transform differently, since the one is determined by
the requirement of leaving the energy-dependent speed of light observer-independent, whereas the other is
determined by the crossing of worldlines of particle for which all {\sc DSR}-effects are negligible. As a consequence, the delay will in some reference frames be larger than or of the same order as
the uncertainty. 

To see this, let us boost into a reference frame with $v=1-\epsilon$, such that $\gamma \approx 1/\sqrt{2 \epsilon}$.
The inequality that needs to be fulfilled to hide the delay is then 
\beqn
\left|\Delta T' - t_a'\right| &\ll& \left |\Delta t' \right| \\
\Leftrightarrow \left|\epsilon - \sqrt{\frac{2}{\epsilon}}\right| &\ll& \epsilon \quad,
\label{see}
\eeqn
which is clearly violated without even requiring extreme boosts.
To put in some numbers, consider an observer in rest with the electron with $\epsilon = 10^{-3}$, and $\gamma \approx 20$.
We then have
\beqn
\left|\Delta T' - t_a' \right| \approx 10^4 \Delta t' \quad.
\eeqn  

Similarly, if we boost into the other direction $v=-1+\epsilon$, the requirement to hide the delay
takes the form
\beqn
\left|\Delta T' - t_a' \right| &\ll& \left |\Delta t' \right| \\
\Leftrightarrow \left|\frac{2}{\epsilon} - \sqrt{\frac{\epsilon}{ 2}} \right| &\ll& \frac{2}{\epsilon} \quad,
\eeqn
which is also clearly violated. Though in this case the delay does not actually get much larger than the uncertainty, they
 both approach the same value. 
We would then be comparing the probability of interaction at
the center of the wave-packet with one at a distance comparable to its width. In this case then the probability of interaction, if we consider a Gaussian wave-packet, had fallen by a
factor of order one. 
Thus, in some reference frames the particles would be able to interact inside the box with some probability 
(depending on the cross-section), whereas in other frames they would only interact in a fraction of these
cases,  in conflict with observer independence.  
This would require several
photons to get a proper statistic, but it is a difference in probability that is feasible to measure within
the lifetime of the universe, and thus is still in conflict with observer-independence The advantage of boosting to a velocity in the opposite direction as the
photon is that the delay itself does not also decrease. 

Let us mention again that we have considered here 
a photon whose approximate uncertainty in momentum space is at emission comparable to the mean value, which is quite badly localized. If the
photon's momentum had instead an uncertainty of $\approx 100$~MeV only, then the mismatch in timescales
was by two orders of magnitude larger. 

We have here assumed that it is appropriate to use the normal Lorentz-boosts to calculate the time span $t'_a$, but
to what precision do we know these? The transformation behavior under 
Lorentz-boosts has been tested to high precision in particle collisions where boosts from the
center of mass system to the laboratory restframe are constantly used. For the time-dilatation in
particular, the decay-time of muons is known 
to transform as $\Delta t' = \gamma \Delta t$ up to a $\gamma$-factor of 30 to a precision 
of one per mille \cite{Bailey:1977de}. 
Note however that $\gamma =30$ is only marginally larger than in the example we have used.
If the arising mismatch thus was a timescale smaller than the scattering process could test,
then we would not have a problem. We will exploit this later to obtain a bound on the
delay still compatible with experiment.

To further distinguish possible options, let us notice that the latter argument actually
referred to an active Lorentz-boost rather than a passive one. An active boost is needed
to describe in our coordinate system properties of the same physical system at different
relative velocities, such as the muons at different rapidity. A passive boost on the 
other hand is used to describe the same physical system as seen from two observers
at different velocities, such as the box in the Earth frame and the  satellite frame. 
In Special Relativity, both boosts are identical (rspt. the one is the inverse of the
other). Due to the human body commonly being in very slow motion compared to elementary 
particles, experimental tests for passive boosts are very limited. In the limit of
small boosts where we can test both, they agree and confirm Special Relativity. Otherwise
we would have to take great care which boost we should be using to describe
signals from {\sc GPS} satellites or read out spectra of atoms in motion \cite{Reinhardt:2007zz}.

We are thus lead to consider the option that
the active boost describing the fast moving muon is not identical with a passive
boost that would be needed to describe the muon/electron from a reference frame at such a
high boost. That would then mean a muon in rest in our reference frame does not
appear to a fast moving observer as the fast moving muon to us. To be concrete,
while the muon's lifetime might be enhanced to $\Delta T' = \gamma_{\rm active} \Delta T$ for
us when we accelerate it, the alien-observer at high $\gamma$ might see our
muon in rest decaying with $\Delta T' = \gamma_{\rm passive} \Delta T$, where 
$\gamma_{\rm passive} \approx 1- v$, such that the box-problem caused by
the different transformation behaviors would be avoided.  That however is either in disagreement
with observer independence or with experiment, which can be seen as follows.

Consider an ultra-high energetic proton that hits
our detector. Are we supposed to describe it by applying an active boost to
protons in rest on Earth, or are we supposed to describe it by a passive boost,
assuming that we should instead transform our coordinate system to that of
the proton? The only way to answer this question is to decide whether
or not the proton has been ``actively'' boosted. But this boost would necessarily
be a boost relative to something. We might for example be tempted to call the
proton actively boosted because it moves fast relative to us or the cosmic microwave background, but
that notion depends on the presence of a preferred frame.
In the case of our
box-problem the question comes down to which reference frame is the right one
to decide whether or not the electron interacts with the slow moving photon
inside the box (with some probability), and why that particular frame
was the right one to pick.

Alternatively, we
could try to find out whether the particle we aim to describe has ever been accelerated 
after its formation. Since acceleration is an absolute notion, the particle's initial 
restframe could then hold as a reference frame to define further active boosts without
singling out a globally preferred frame. Leaving aside the problem of defining a
restframe for massless particles, this would mean the boost we needed to describe a 
particle depended on the previous history of the particle. In particular this would
mean properties of particles produced at high rapidity in a collision would have to be
transformed into the lab frame by a passive boost. This boost would in high energy collisions have to
differ by many orders of magnitude from the standard Lorentz-transformation, a
modification we would long have seen. But in addition, this would mean that the
muon-decay actually does probe passive rather than active boosts and thus provides
the constraint we were using.

To summarize this argument, we have seen that an increased quantum mechanical
uncertainty $\Delta t$ that scales with the delay between the high- and low-energetic
photon $\Delta T$ cannot in all reference frames bridge the distance the photon is
lagging behind the electron when we use a normal Lorentz-boost. And that even though
we have used an at emission very badly localized photon already. Active boosts
have been tested up to the necessary precision such that a delay of $\Delta T$
of the order seconds would result in a conflict with observer-independence. 
If passive boosts were different
from active boosts, this would necessitate the introduction of a preferred frame
and thus disagree with our aim to preserve observer independence. Either way
we turn it, quantum mechanics does not solve the box-problem. We will thus
in the following section draw consequences.

It is worthwhile to note however that adding quantum mechanical uncertainty does
solve the box problem, version 1.0, discussed in section \ref{1.0}. This is because,
as previously noted, {\sc DSR} generically also implies a modification of the maximally possible
localization due to an energy dependence of Planck's constant. Take for example the dispersion
relation \cite{Magueijo:2002am}:
\beqn
\frac{E^2}{\left( 1+ E/m_{\rm Pl} \right)^2} = p^2 \quad.
\eeqn
It has the property of setting a maximal possible value for the momentum, 
$p = m_{\rm Pl}$, which is only reached for $E \to \infty$. In this case the 
energy-dependent speed of light and Planck's constant are \cite{Hossenfelder:2006rr}:
\beqn
\tilde c (E) = \frac{1}{{1 + E/m_{\rm Pl}}}\quad,\quad \tilde \hbar(E) = {1+E/m_{\rm Pl}} \quad.
\eeqn
Thus, while the speed of light goes to zero, Planck's constant goes to infinity. For the photon in
rest in the box this would
result in an infinite position uncertainty, such that neither observer could plausibly
say whether the particle is inside the box or not.

\section{Bounds}
\label{bounds}

What if we tried to live with the electron scattering off the photon 10
meters outside the detector? This would require
the cross-section for Coulomb-scattering in the satellite frame to be dramatically
different from what we have measured in the Earth frame. In the Earth frame, this
scattering process probes a typical distance inverse to the center of mass
energy of the scattering particles. In the satellite frame, the cross-section
must be the same for the distance the photon is lagging behind the electron.
This cross-section might not indeed have been measured in any satellite, but this is unnecessary because 
if it was different from that in our Earth frame this would be incompatible with observer-independence. 

The logic of the here presented argument is as follows. If there was an energy-dependent
speed of light that resulted in the 10~GeV photon arriving about 1 second later than
the low-energetic photon, then the requirement of observer-independence implies 
violations of locality that are incompatible with previously made experiments. 
Note that it is not necessary to actually perform the experiment as in the
setup explained in the previous sections since observer-independence means we can
rely on cross-sections previously measured on Earth. In that sense, the experiment has
already been done. 
The setup has only been added to 
make clear that the effect is not in practice undetectable and thus cannot be
discarded as a philosophical speculation. To then resolve the disagreement, 
we either have to give up observer-independence, which would mean we are not
talking about {\sc DSR} any longer, or, if we want to stick with {\sc DSR}, the
violations of locality should be small enough to not be in conflict with any already made
experiment. 

This means one can use the excellent knowledge of {\sc QED} processes to 
constrain the possibility of there
being such a {\sc DSR} modification by requiring the resulting mismatch in arrival
times not to result in any conflict with cross-sections we have measured. 

Let us first consider the case where there is no {\sc DSR}-modification of the
quantum mechanical uncertainty. The distance $L =$ some Gpc is as high as we can plausibly get in
our universe, and the $10$ GeV photon is as high as we have reliable observational
data from particles traveling that far. The center of mass energy of the electron
and the high energetic photon is $\sqrt{s} \approx 15$~MeV. The process
thus probes distances of $\approx 10$~fm. If the photon and
the electron were in the satellite frame closer already than the distance their 
scattering process probes, we would not have a problem. 
Requiring $|\Delta T' -t'_a| < 10$~fm leads to a bound on the delay between the low
and high energetic photon of
\beqn
\Delta T < 10^{-17} {\rm s} \quad,
\eeqn
in order for there not to be any conflict with known particle physics.
If we reinsert the $\alpha$ that we set to one from
Eq. (\ref{alpha}), we can write the bound as $\alpha < 10^{-18}$. This is what
we find from the requirement that there be no problem in the satellite frame in
the case without an additional dispersion of the photon's wave-packet. With such a dispersion,
there is no problem in the satellite frame.

However, according to our argumentation in the previous section we can trust 
Lorentz-boosts up to $\gamma \approx 30$. Using
this boost increases the  mismatch to $|\Delta T' - t_a'| \approx 80 \Delta T$, and
the requirement that it be unobservable with presently tested {\sc QED} precision amounts
to
\beqn
\Delta T < 10^{-23} {\rm s} \quad,
\eeqn 
or $\alpha < 10^{-24}$. Note that this does take into account a possible {\sc DSR}-modification
of quantum mechanics already, and thus covers both cases, the one with and without
spread of the wave-packet. However, since the ratio $E_\gamma/m_{\rm Pl}$ is approx $10^{-18}$,
 present-day observations do already rule out any 
first order modification in the speed of light, and come
indeed close to testing a second order modification. The here offered
analysis however depends on the scaling in Eq. (\ref{DeltaT}) and thus applies only
for modifications linear in the energy.

It is quite possible that the energies we have chosen and the setup we have used
do not yield the tightest constraints possible. One could for example have used a photon 
scattering off another photon or more complicated scattering processes 
involving neutrinos or other light elementary particles, or have the electron
be emitted from a different source such that the center of mass energy is higher. 
We will not examine all of these cases here, but it seems feasible to get the 
bound another one or two orders of magnitude stronger. Even stronger constraints
might arise from considering high energetic scattering processes in the early universe.

\section{Discussion}
\label{disc}

Let us now see whether there are other options to save {\sc DSR} in face of the box problem. 
First we notice 
that the problem evidently stems
from the transformation behavior of $\Delta T$ in Eq. (\ref{dt}). This behavior is
a direct consequence of requiring the energy-dependent speed of light $\tilde c$ to be
observer-independent, together with applying a normal, passive, Lorentz-transformation
to convert the distance $L$ into the satellite restframe. Now if one would use a 
modified Lorentz-transformation also on the coordinates, a transformation
depending on the energy of the photon, then $\Delta T$ could indeed transform properly 
and both particles would meet also in the satellite frame. This would require that
the transformation on the distance $L$ was modified such that it converted
the troublesome transformation behavior of $\Delta T$ back into a normal 
Lorentz-transformation. Then, all observers would agree on their observation.
  
The consequence of that would be that the distance between any two objects would depend on the energy of a photon
that happened to propagate between them, an idea that is hard to make sense of. But
even if one wants to swallow this, the result would just be that the distance between
the {\sc GRB} and the detector was energy-dependent such that it got shortened in the
right amount to allow the slower photon to arrive in time together with the electron. 
That however meant of course the speed of the photon would not depend on its energy. 
The confusion here stems from having defined a speed from the dispersion relation 
without that speed a priori having any meaning in position space. Thus, this
possibility does indeed solve the box-problem, but just reaffirms that
observer-independence requires the speed of light to be constant.

Or, one might want to argue that maybe in the satellite frame the both photons
were not emitted at the same time, such that still the electron could arrive
together with the high energetic photon. This however just pushes the bump around under
the carpet by moving the mismatch in the timescales in the satellite frame 
away from the detector and towards the source. One could easily construct
another example where the mismatch at the source had macroscopic consequences.
This therefore does not help solving the problem.

Another option would be to exploit that the problem arises from the same
fact that made the time-delay of the photon observable in the first line: 
the long distance traveled. One could thus demand the cross-section
to depend on the history of the photon, such that it was only the long-traveled
photons that required strong modifications on {\sc QED} cross-sections. Basically,
this would mean that any particle's cross-section was dependent on the
particle's history. This is unappealing, but worse, then
cross-sections had to be modified for all ultra-high energetic
particles that have travelled long distances, and there is so far no 
indication for that. In particular, since interstellar space is not actually empty, a 
large increase in the photon-photon cross-section 
would not allow the high energetic photons to arrive on Earth at all. 

Then, finally, one could try to accept that the electron just does not scatter
off the photon. This would mean that the macroscopic history an observer
sees depended on his relative velocity. This would certainly have made stays in space stations
much more interesting. 

Let us point out that the box-problem does not exist in theories that break rather
than deform Lorentz-invariance. The reason is that in the case Lorentz-invariance is broken, the speed
of the high energetic photon is not an observer-independent function of the
energy. Instead, the relations (\ref{disp}) and (\ref{cofe}) only hold in one
particular frame, and in all other frames they contain the velocity relative
to that particular frame. There are however strong constraints on the breaking
of Lorentz-invariance already from many other observations, see e.g. \cite{Maccione:2007yc} and
references therein.

We started with the motivation that the requirement of the Planck energy
being observer-independent seems to necessitate a modification of Lorentz-invariance
that can result in an energy-dependent speed of light.
This energy-dependent speed of light has then lead us to violations of locality 
that are hard to reconcile
with experiment. That {\sc DSR} implies a frame-dependent meaning of what
is ``near'' was mentioned already in
\cite{AmelinoCamelia:2002vy}. Serious conceptual problems arising from
this were pointed out in \cite{Schutzhold:2003yp,Hossenfelder:2006rr},
and here we demonstrated a conflict with experiment to very
high precision. 

It has however been argued in \cite{Hossenfelder:2006cw} that
the requirement of the Planck scale being observer-independent does not
necessitate it to be an invariant of Lorentz-boosts, since the result of
such a boost does not itself constitute an observation. It is sufficient 
that experiments made are in agreement over that scale. In
particular if the Planck length plays the role of a fundamentally minimal
length no process should be able to resolve shorter distances. This does
require a modification of interactions in quantum field theory at
very high center-of-mass energies and small impact parameters, but it
does not necessitate a modification of Lorentz-boosts for free particles. 
In this case, the speed of light remains constant and the box
is not a problem. 

\section{Conclusion}

We have studied the consequences of requiring an energy-dependent and
observer-indepen\-dent speed of light in Deformed Special Relativity. We have
shown it to result in an observer-dependent notion of what constitutes
the same space-time event and thus were lead to consider violations
of locality arising from such a transformation behavior. Using the concrete example 
of a highly energetic photon emitted
from a distant gamma ray burst, we have shown that these violations of
locality would be in conflict with already measured elementary 
particle interactions if the energy dependence was of first order in the energy
over the Planck mass. This in turn was used to derive a bound on the
still possible modifications in the speed of light, which is 22
orders of magnitude stronger than previous bounds that were obtained from
direct measurements of delays induced by the energy-dependence. 
This new bound rules out modification to first order in the 
energy over the Planck mass.

\section*{Acknowledgements}

I want to thank Giovanni Amelino-Camelia, Stefan Scherer, and Lee Smolin for helpful
comments.

\end{document}